\documentclass[a4paper,11pt]{article}
\usepackage{pos}
\usepackage{booktabs}
\usepackage{amsmath}
\usepackage{array}
\usepackage{graphicx}

\title{New results from electron-positron heavy-flavor factories}
\ShortTitle{New results from e$^+$e$^-$ heavy-flavor factories}

\author*[a]{Xiaodong Shi}
\affiliation[a]{Institute of High Energy Physics, Chinese Academy of Sciences,\\
  Beijing 100049, China}

\emailAdd{shixd@ihep.ac.cn}

\abstract{The electron-positron heavy-flavor factories --- BaBar, Belle, Belle~II, and
BESIII --- continue to deliver a broad and sensitive physics program.
I report selected highlights presented at ICHEP 2026, covering neutral- and charged-current $B$
processes, tests of lepton-flavor universality in $b\to c\ell\nu$ transitions, precision
measurement of Cabibbo--Kobayashi--Maskawa matrix elements, and hadron spectroscopy.
}

\FullConference{43rd International Conference on High Energy Physics (ICHEP 2026)\\
30 July -- 5 August 2026\\
Natal, Brazil}

\begin{document}
\maketitle

\section{Introduction}

The Standard Model (SM) describes the elementary particles and their
interactions with remarkable success. However, it remains incomplete:
it does not explain the matter-antimatter asymmetry of the Universe,
provides no viable candidate for dark matter, and leaves open fundamental
questions concerning the origin of flavor and fermion mass hierarchies,
the strong CP problem, and the stability of the electroweak scale
\cite{PDG2026}. These open questions provide compelling motivation for
physics beyond the SM~\cite{PDG2026,Altmannshofer:2025}.

Two complementary strategies are pursued at colliders: direct searches
for on-shell production of new particles at the energy frontier, and
indirect searches at the flavor frontier, where new physics may enter
through quantum loops and manifest itself as deviations in precision
observables \cite{Altmannshofer:2025}.
The latter can probe new-physics scales $\Lambda_{\rm NP}$ far above the energies
directly accessible. Precision flavor measurements have historically provided indirect
evidence for particles before their direct observation, most notably the charm quark and the top quark. 
A solid understanding of quantum chromodynamics (QCD) is essential, however, because
non-perturbative effects must be controlled at a level commensurate with the experimental
precision.

Electron-positron heavy-flavor factories are particularly well suited to search for new physics. Their clean
environment enables excellent reconstruction of neutral particles, while the precisely known
initial-state four-momentum and constrained event kinematics provide powerful background
rejection and access to final states with missing energy.
At the $\Upsilon(4S)$, the $B^0\bar B^0$ pair is produced in a coherent, quantum-entangled state,
providing additional information for time-dependent measurements. These advantages make
electron-positron heavy-flavor factories complementary to the Large Hadron Collider (LHC).

\section{The electron-positron heavy-flavor factories and their datasets}
\label{sec:factories}

The first-generation $B$ factories, BaBar at PEP-II and Belle at KEKB, collected
$0.53~\mathrm{ab}^{-1}$ and $1.04~\mathrm{ab}^{-1}$, respectively, with most of these data
taken at the $\Upsilon(4S)$ resonance. The second-generation Belle~II at SuperKEKB~\cite{ref:belle2tdr} has accumulated
$0.91~\mathrm{ab}^{-1}$ since 2019 and surpasses Belle's 
$\Upsilon(4S)$ data sample as of 2026. BESIII at BEPCII~\cite{ref:bes3future} is a $\tau$-charm factory that has collected
about $50~\mathrm{fb}^{-1}$ since 2009 over center-of-mass energies from 1.84 to
$4.95~\mathrm{GeV}$.



\section{$B$ anomalies}
\label{sec:banomalies}

The flavor anomalies discussed here fall into two categories: neutral-current processes
mediated by the $b\to s\ell^+\ell^-$ and $b\to s\nu\bar\nu$ transitions, which occur only at loop level in the SM and are therefore sensitive to new physics,
and charged-current $b\to c\ell\nu$ transitions, which test whether the electroweak
interaction couples with equal strength to the three lepton generations.

\subsection{Neutral-current processes}

A representative example of the $b\to s\nu\bar\nu$ transition is $B^+\to K^+\nu\bar\nu$.
With only a charged kaon visible in the final state of the signal $B$ decay, this channel is experimentally
challenging. 
Belle~II has reported evidence for this decay at the $3.5\sigma$
level~\cite{ref:knuunu}, with the measured branching fraction
$\mathcal{B}(B^+\to K^+\nu\bar\nu) = (2.3 \pm 0.5\,(\mathrm{stat})\,^{+0.5}_{-0.4}\,(\mathrm{syst}))\times 10^{-5}$. This value lies $2.7\sigma$ above the SM expectation. 
Its interpretation as evidence for new physics is nevertheless model-dependent.
Because the selection efficiency depends on the
dineutrino squared mass $q^2$, new physics that alters the $q^2$ spectrum cannot be tested by
rescaling the branching fraction;
Belle~II has therefore published the full likelihood together with the joint number density of simulated Standard Model signal events.
As an application, the result is reinterpreted in the weak
effective theory, in which the differential branching fraction depends on vector, scalar, and
tensor Wilson coefficients. A Bayesian fit finds that the data prefer an enhanced vector
contribution and mildly favour a nonzero tensor contribution, indicating that the SM $q^2$ shape
is not the best description of the data~\cite{ref:wilson}.

Related channels can be probed using similar missing-energy signatures. The first search for inclusive
$B\to X_s\nu\bar\nu$ decays at Belle~II, in which the strange hadronic system $X_s$ is explicitly reconstructed,
yields
$\mathcal{B}(B\to X_s\nu\bar\nu) < 3.3\times 10^{-4}$ at the 90\% confidence
level~\cite{ref:xsnn}, while the corresponding measurement of $B\to X_s\ell^+\ell^-$ gives
$\mathcal{B}(B\to X_s\ell^+\ell^-) = (3.91 \pm 0.49\,(\mathrm{stat})\,^{+0.38}_{-0.37}\,(\mathrm{syst}))\times 10^{-6}$
and is being prepared for publication.

The $b\to s\tau^+\tau^-$ transition probes new physics that couples preferentially to the
third generation, and the Belle and Belle~II collaborations have performed the first search for
$B^0\to K^0_S\tau^+\tau^-$ using the full Belle and Belle~II Run~1 $\Upsilon(4S)$ datasets, with
the non-signal $B$ reconstructed through hadronic tagging~\cite{ref:ktatau}. No evidence is
observed, and the first upper limit on this decay is
$\mathcal{B}(B^0\to K^0_S\tau^+\tau^-) < 8.3\times 10^{-4}$ at the 90\% confidence level,
about four orders of magnitude above the SM prediction
$\mathcal{B}(B^0\to K^0_S\tau^+\tau^-) \simeq 0.8\times 10^{-7}$~\cite{ref:hpqcd}.
Combining with a companion search for the isospin partner $B^+\to K^+\tau^+\tau^-$, which sets
$\mathcal{B}(B^+\to K^+\tau^+\tau^-) < 5.6\times 10^{-4}$ at the 90\% confidence
level~\cite{ref:kptautau}, Belle~II determines $\mathcal{B}(B\to K\tau^+\tau^-) < 5.4\times 10^{-4}$ at the 90\% confidence
level~\cite{ref:ktatau}, constraining scalar and pseudoscalar contributions to $b\to s\tau^+\tau^-$.

\subsection{Lepton-flavor universality: $R(D)$ and $R(D^*)$}
\label{sec:rd}

In the SM, the electroweak gauge bosons couple universally to electrons, muons, and
tau leptons. Deviations from this universality can arise from interactions beyond the SM.
Semileptonic $B$ decays provide a theoretically clean test because many hadronic uncertainties
partially cancel in the ratios
\begin{equation}
R(D^{(*)}) = \frac{\mathcal{B}(B\to D^{(*)}\tau^-\bar\nu_\tau)}
                  {\mathcal{B}(B\to D^{(*)}\ell^-\bar\nu_\ell)},
\label{eq:rd}
\end{equation}
where $\ell$ denotes an electron or a muon. 
The other $B$ meson must be reconstructed to constrain the kinematics, either through a hadronic tag (such as
$D^{(*)}\pi$) or a semileptonic tag (such as $D\ell\nu$). The former generally offers higher purity and better kinematic resolution, whereas the latter
provides higher efficiency; the two strategies also probe statistically independent samples when
constructed from disjoint tag selections.

\begin{figure}[b]
\centering
\includegraphics[width=0.85\textwidth]{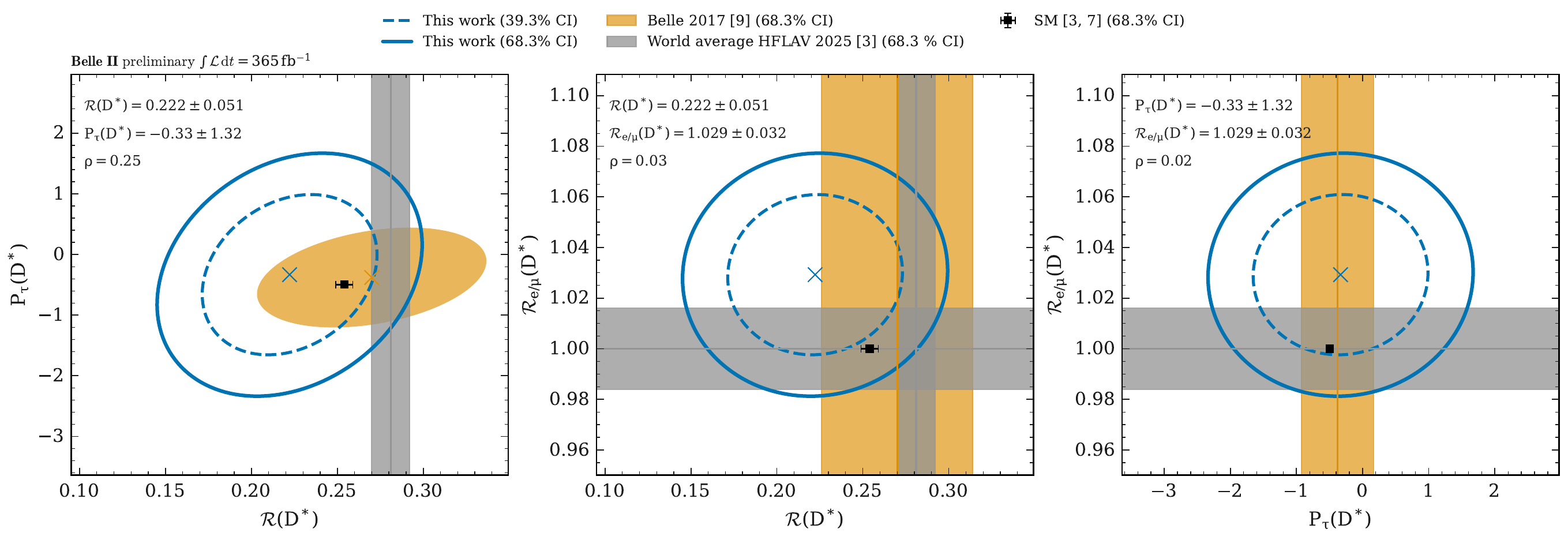}
\caption{Confidence regions in the $R(D^*)$--$P_\tau(D^*)$, $R(D^*)$--$R_{e/\mu}(D^*)$, and
$P_\tau(D^*)$--$R_{e/\mu}(D^*)$ planes. The result of this measurement is
compared to the $R(D^*)$ world average~\cite{ref:hflav} and the previous $P_\tau(D^*)$
measurement.}
\label{fig:rdstar}
\end{figure}

Belle~II has presented a new simultaneous measurement of $R(D^*)$ and the $\tau$-lepton
polarization $P_\tau(D^*)$ using $365.3 \pm 1.7~\mathrm{fb}^{-1}$ of $\Upsilon(4S)$ data, with a
hadronic tag and the $\tau$ reconstructed in the helicity-sensitive modes $\tau^-\to\pi^-\nu_\tau$
and $\tau^-\to\rho^-\nu_\tau$. The two observables are extracted together
to distinguish between different new-physics scenarios. The simultaneous fit gives
\begin{equation}
\begin{aligned}
R(D^*) &= 0.222 \pm 0.041\,(\mathrm{stat}) \pm 0.030\,(\mathrm{syst}),\\
P_\tau(D^*) &= -0.33 \pm 1.15\,(\mathrm{stat}) \pm 0.65\,(\mathrm{syst}),\\
R_{e/\mu}(D^*) &= 1.029 \pm 0.029\,(\mathrm{stat}) \pm 0.014\,(\mathrm{syst}),
\end{aligned}
\label{eq:belle2rdstar}
\end{equation}
where $R_{e/\mu}(D^*) = \mathcal{B}(B\to D^*e^-\bar\nu_e)/\mathcal{B}(B\to D^*\mu^-\bar\nu_\mu)$
tests universality between the two light charged leptons. The fit yields a correlation
$\rho = 0.25$ between $R(D^*)$ and $P_\tau(D^*)$ and a goodness-of-fit $p$-value of 12.2\%; the
confidence regions are shown in Fig.~\ref{fig:rdstar}. Relative to the
Standard Model predictions $R(D^*) = 0.254 \pm 0.005$ and $P_\tau(D^*) = -0.497 \pm
0.007$~\cite{ref:fajfersm,ref:bernlochner,ref:bigi}, with $R_{e/\mu}(D^*)$ expected to be close to
unity, the three results correspond to pulls of $0.62\sigma$, $0.12\sigma$, and $0.92\sigma$,
respectively, and are consistent with the SM within $0.68\sigma$ in combination.

A new BaBar measurement of $R(D)$ and $R(D^*)$, based on a semileptonic-tag sample and
therefore statistically independent of its earlier hadronic-tag result, has also been
presented. The preliminary result is: 
\begin{equation}
\begin{aligned}
R(D) &= 0.316 \pm 0.062\,(\mathrm{stat}) \pm 0.019\,(\mathrm{syst}),\\
R(D^*) &= 0.226 \pm 0.022\,(\mathrm{stat}) \pm 0.012\,(\mathrm{syst}),\\
\end{aligned}
\label{eq:Babarrdstar}
\end{equation}
which is consistent with the SM at the
$1.5\sigma$ level; some systematic uncertainties and consistency checks are still ongoing.

\begin{figure}[b]
\centering
\includegraphics[width=0.50\textwidth]{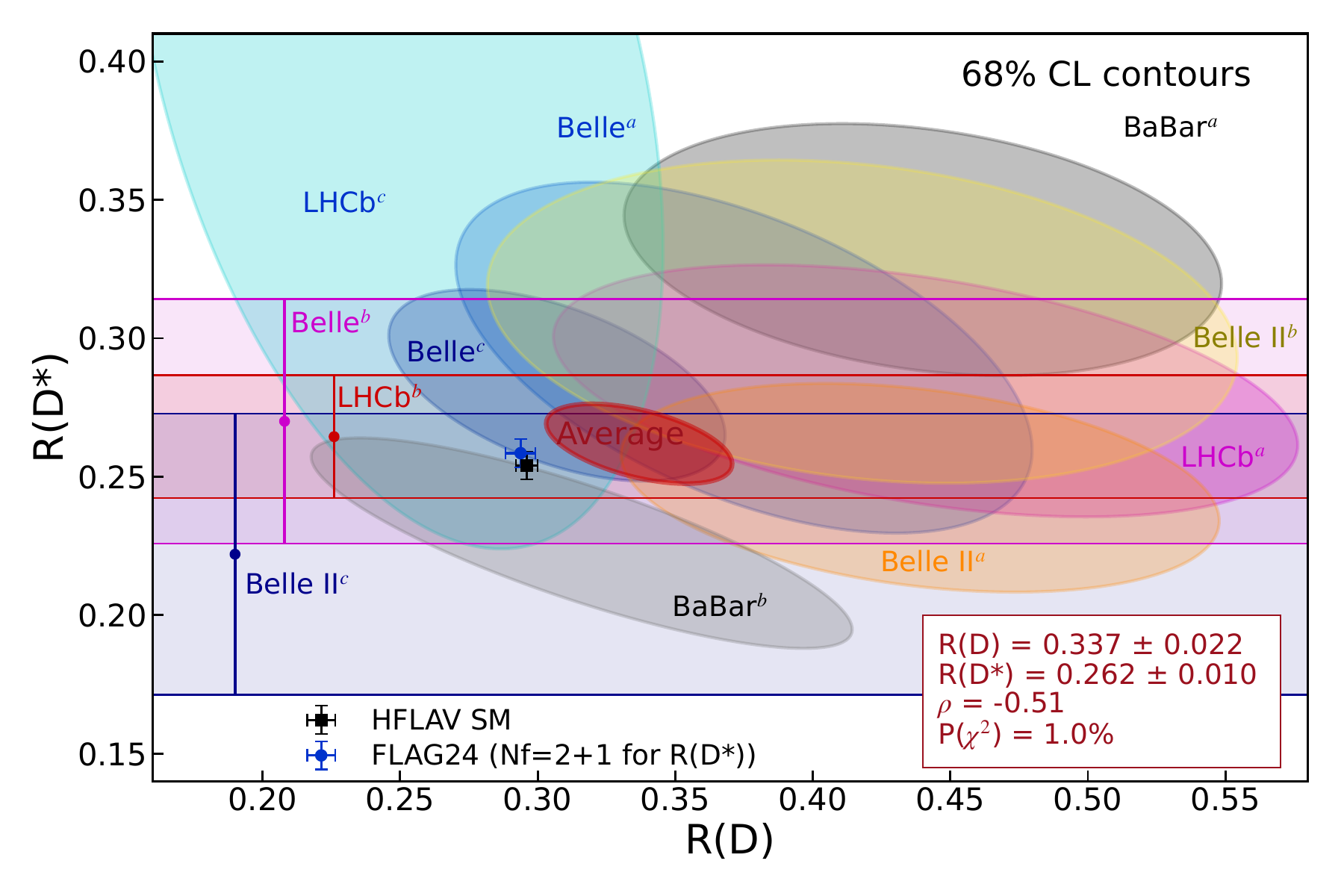}
\caption{The $R(D)$--$R(D^*)$ plane, produced by the HFLAV collaboration, showing the 68\% confidence
contours of the individual measurements together with the world average~\cite{ref:hflav} and
the Standard Model prediction; the combination displayed is the preliminary ICHEP 2026
update.}
\label{fig:rd}
\end{figure}

The individual measurements of $R(D)$ and $R(D^*)$ are collected in Fig.~\ref{fig:rd}, from HFLAV, together with the world
average~\cite{ref:hflav}. Their combination gives $R(D) = 0.337 \pm 0.022$ and
$R(D^*) = 0.262 \pm 0.010$, with $\rho = -0.51$ and a small $p$-value of 1.0\%.
Compared to SM predictions, the discrepancy is at the level of
approximately $2\sigma$, reduced from roughly $3\sigma$ before these two preliminary results were
incorporated into the average. 
\footnote{HFLAV group also provide a combination result w/o BaBar's preliminary result, which has a bit higher tension to SM predction.~\cite{ref:hflav}.}
With more data from both Belle~II and LHCb, as well as better control of background events, this
anomaly will continue to evolve.

\section{CKM metrology}
\label{sec:ckm}

The Cabibbo--Kobayashi--Maskawa (CKM) matrix governs quark flavor-changing transitions
and determines the size of CP violation in the quark sector~\cite{ref:km}. In the SM it is fully described
by four parameters, and its unitarity provides powerful consistency conditions that can be
tested with many independent measurements. These include the vanishing sums of products of
matrix elements along rows or columns. The familiar example
$V_{ud}^{}V_{ub}^* + V_{cd}^{}V_{cb}^* + V_{td}^{}V_{tb}^* = 0$ is represented geometrically by the
unitarity triangle in the $(\bar\rho,\bar\eta)$ plane. Unitarity also implies normalization
conditions, such as $\sum_i |V_{ij}|^2 = 1$ for each down-type quark index $j$.

\subsection{Towards the apex: the first measurement of $S(\pi^0\pi^0)$}

\begin{figure}[b]
\centering
\includegraphics[width=0.38\textwidth]{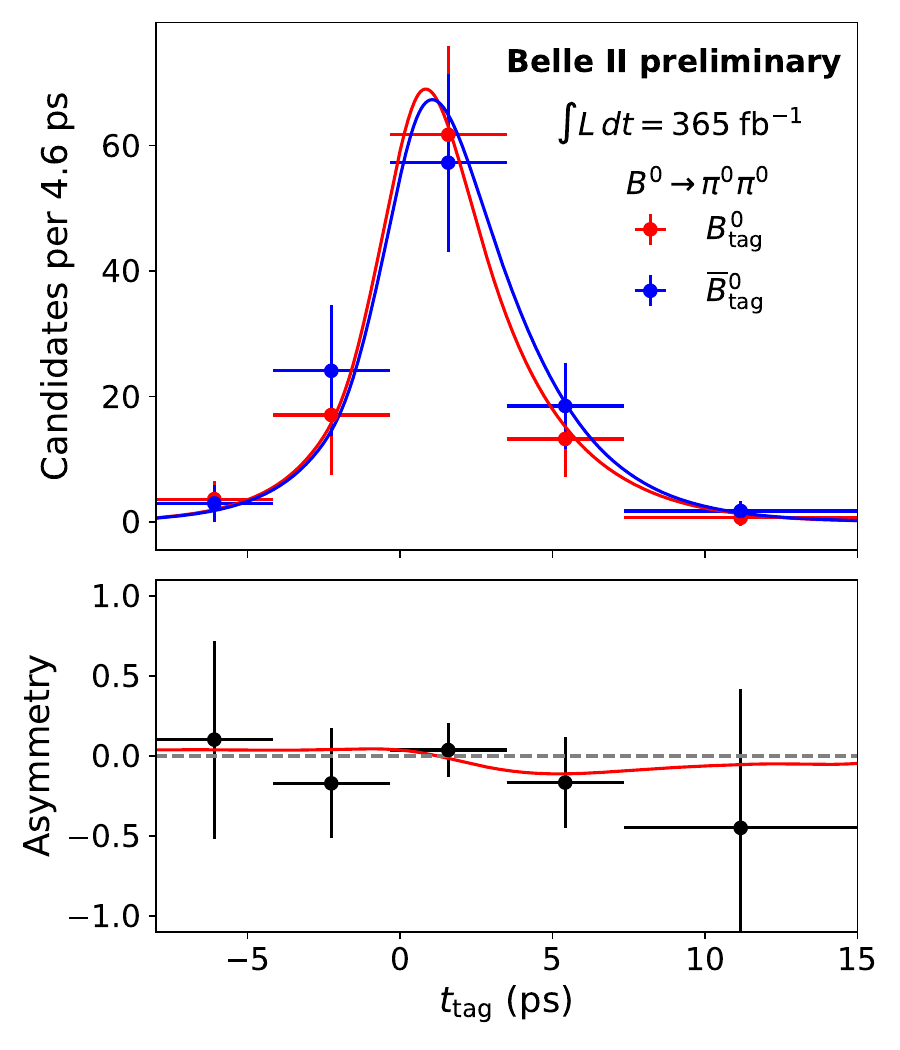}\hfill
\includegraphics[width=0.572\textwidth]{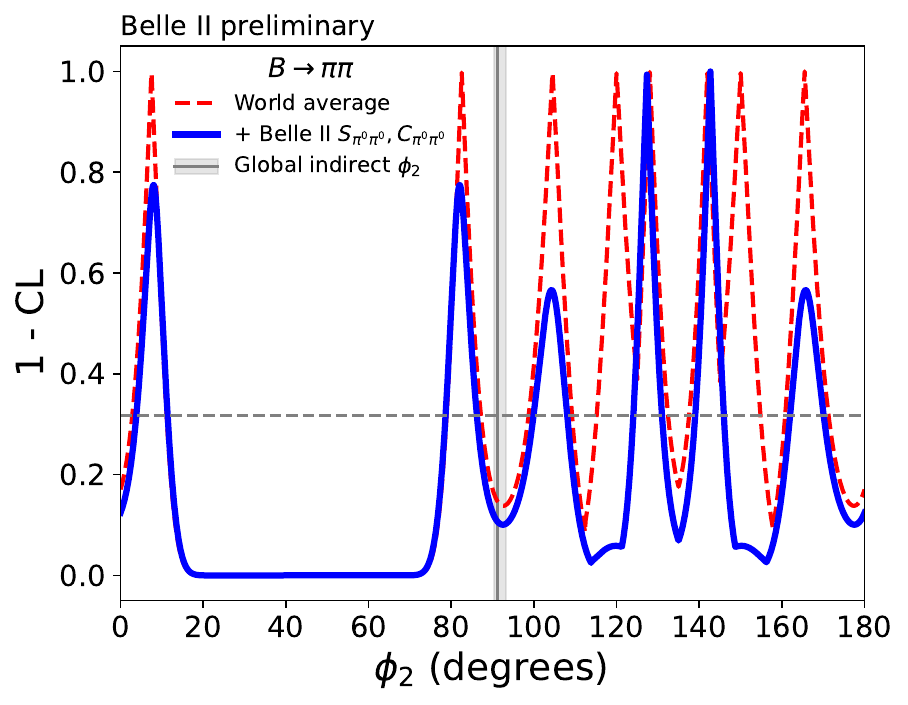}
\caption{(a) Distribution of the tag-side $B$ decay time $t_{\rm tag}$ for background-subtracted
$B^0\to\pi^0\pi^0$ candidates at Belle~II, from which $S(\pi^0\pi^0)$ and $C(\pi^0\pi^0)$
are extracted. (b) Values of $1-{\rm CL}$ as a function of $\phi_2$ determined from $B\to\pi\pi$
decays (dashed red) before and (solid blue) after including the $S(\pi^0\pi^0)$ and
$C(\pi^0\pi^0)$ measurements reported here; the vertical line indicates the indirect value of
$\phi_2$ from the global CKM fit and the horizontal dashed line the 68.3\% confidence
level.}
\label{fig:spp}
\end{figure}

The angle $\phi_2$ (also denoted $\alpha$) remains among the less precisely determined
angles of the unitarity triangle. Experimentally, it is extracted from isospin analyses of the
$B\to\pi\pi$ and $B\to\rho\rho$ systems, which require the branching fractions and CP
asymmetries of all relevant charge states as input. A long-standing missing ingredient in
the $B\to\pi\pi$ analysis has been the mixing-induced CP asymmetry $S(\pi^0\pi^0)$.
The conventional method based on the decay-time difference $\Delta t$ requires reconstruction
of the decay vertices of both $B$ mesons, which is extremely challenging for the $B^0\to\pi^0\pi^0$ decay.

Belle~II has overcome this obstacle with a novel method that requires only the vertex of the
tag-side $B$ meson~\cite{ref:tagvertex}. Because the two $B$ mesons produced at the $\Upsilon(4S)$ are in a
coherent entangled state, the tag-side flavor and decay-time information can be used to extract
$S(\pi^0\pi^0)$ without reconstructing the signal-side decay vertex.
The nano-beam scheme of SuperKEKB and the pixel detector at Belle~II enable the tag-side vertex to be
determined precisely.

The measurement uses $170$ signal events selected from $365~\mathrm{fb}^{-1}$ of Belle~II data, with
a flavor-tagging efficiency of 37\% obtained using a graph neural-network tagger~\cite{ref:gnn}. The decay-time
distribution is shown in Fig.~\ref{fig:spp}(a), and Belle~II obtains
\begin{equation}
S(\pi^0\pi^0) = 0.61\,^{+0.75}_{-0.79}\,(\mathrm{stat}) \pm 0.11\,(\mathrm{syst}),
\qquad
C(\pi^0\pi^0) = 0.05 \pm 0.28\,(\mathrm{stat}) \pm 0.07\,(\mathrm{syst}),
\label{eq:spp}
\end{equation}
This is the first measurement of mixing-induced CP violation in this decay~\cite{ref:spp}. Achieving comparable precision with a conventional method would require a data set about twenty times larger. Incorporating this result into the determination of $\phi_2$ from $B\to\pi\pi$, shown in Fig.~\ref{fig:spp}(b), narrows
the $1\sigma$ interval by 40\% and reduces the ambiguity in the extracted value~\cite{ref:ckmfitter}.

\subsection{$|V_{cs}|$ and $|V_{cd}|$ from charm semileptonic decays}

The CKM elements involving the charm quark can be determined from semileptonic $D$ decays,
with hadronic form factors providing the essential non-perturbative input. BESIII has exploited
the world's largest samples of $D\bar D$ pairs produced at threshold~\cite{ref:bes3future},
corresponding to $20.3~\mathrm{fb}^{-1}$. A study of $D^0\to K^-\ell^+\nu_\ell$ and $D^+\to\bar K^0\ell^+\nu_\ell$ gives
$f_+(0)|V_{cs}| = 0.7183 \pm 0.0007\,(\mathrm{stat}) \pm 0.0014\,(\mathrm{syst})$.
Combined with the form factor from lattice QCD, this yields~\cite{ref:vcs}
\begin{equation}
|V_{cs}| = 0.9639 \pm 0.0009\,(\mathrm{stat}) \pm 0.0018\,(\mathrm{syst}) \pm 0.0040\,(\mathrm{LQCD}),
\label{eq:vcs}
\end{equation}
a total precision of about 0.5\%, with the uncertainty dominated by the lattice-QCD form
factor rather than by the experimental measurement.
A companion analysis provides the first experimental constraint on the scalar current in the
$c\to s\ell^+\nu_\ell$ transition~\cite{ref:scalark}. An analogous study of
$D^0\to\pi^-\ell^+\nu_\ell$ and $D^+\to\pi^0\ell^+\nu_\ell$ gives
$f^{D\to\pi}_+(0)|V_{cd}| = 0.1425 \pm 0.0005\,(\mathrm{stat}) \pm 0.0003\,(\mathrm{syst})$
and hence~\cite{ref:vcd}
\begin{equation}
|V_{cd}| = 0.2262 \pm 0.0008\,(\mathrm{stat}) \pm 0.0005\,(\mathrm{syst}) \pm 0.0018\,(\mathrm{LQCD}),
\label{eq:vcd}
\end{equation}
a total precision of about 0.9\%, again with the uncertainty dominated by the lattice-QCD
form factor. The precision
reach of flavor physics is increasingly limited by non-perturbative QCD inputs~\cite{ref:decdyn}.

\section{Hadron spectroscopy: probing non-perturbative QCD}
\label{sec:spectroscopy}

Non-perturbative QCD is notoriously difficult to calculate, and hadron spectroscopy provides a direct
way to study its spectrum and dynamics. Beam-energy scans at $e^+e^-$ colliders provide a particularly clean environment because the
initial state is precisely defined, allowing exclusive production cross sections and line shapes
to be studied as functions of the center-of-mass energy.

\subsection{The conventional hadrons: first evidence for $\Upsilon_1(1D)$ and $\Upsilon_3(1D)$}

In the quark model, bottomonium states are classified by the orbital angular momentum $L$ and
total spin $S$ of the $b\bar b$ pair. For the $1D$ triplet, $L=2$ and $S=1$, giving
$J^{PC}=1^{--},2^{--},3^{--}$. These states are less well established experimentally than
the corresponding $S$-wave states.
Until now, only the $\Upsilon_2(1D)$ state had been observed. Using
$122~\mathrm{fb}^{-1}$ of Belle data collected at the $\Upsilon(5S)$ resonance
($\sqrt s = 10\,865.8~\mathrm{MeV}$) together with energy-scan data at 18 further
center-of-mass energies between 10.73 and 11.02~GeV, the Belle and Belle~II collaborations
have studied $e^+e^-\to\Upsilon_J(1D)\eta$ and $e^+e^-\to\Upsilon_J(1D)\pi^+\pi^-$~\cite{ref:yd}.
The analysis finds first evidence for the $\Upsilon_1(1D)$ and $\Upsilon_3(1D)$ states at
$4.8\sigma$ and $3.0\sigma$, respectively, while the $\Upsilon_2(1D)$ signal is established
at more than $10\sigma$. The mass of the $\Upsilon_2(1D)$ is
$(10\,167.0 \pm 1.0 \pm 0.2)~\mathrm{MeV}/c^2$, and the splittings within the triplet are $\Delta m_{12} = (11.8 \pm 1.5 \pm 0.4)~\mathrm{MeV}/c^2$ and
$\Delta m_{23} = (7.6 \pm 2.4 \pm 0.6)~\mathrm{MeV}/c^2$. These measurements provide a test of model and lattice-QCD predictions for the spin-dependent
interactions between the heavy quark and antiquark.

\subsection{The unconventional hadrons: the $X(2370)$ and the lightest glueball}

\begin{figure}[b]
\centering
\includegraphics[height=0.274\textheight]{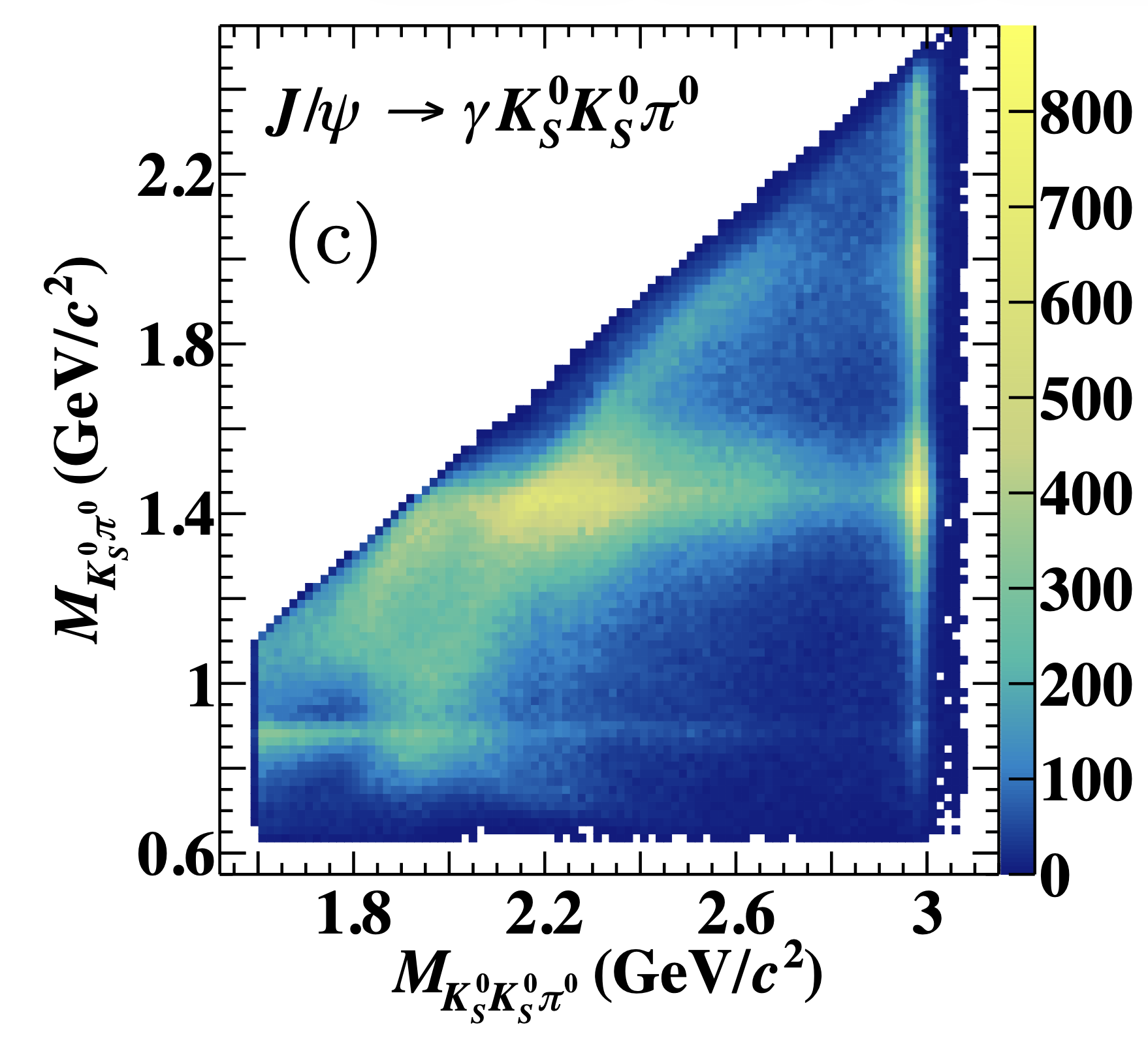}\hfill
\includegraphics[height=0.274\textheight]{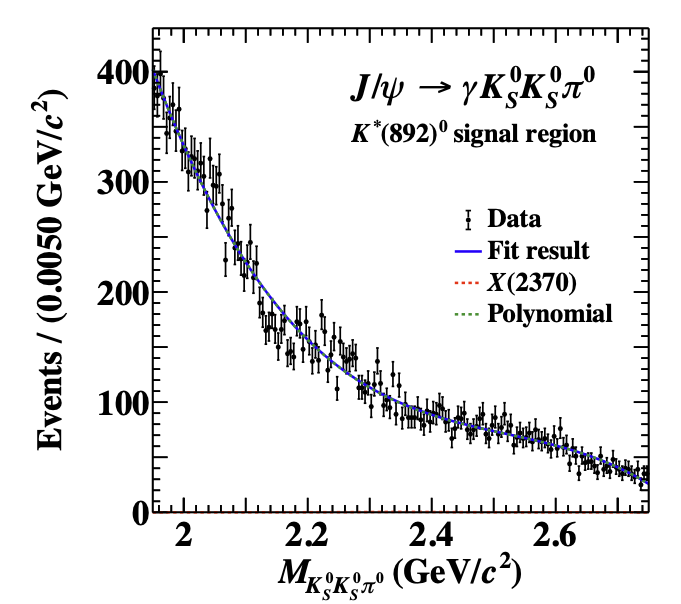}
\caption{(a) The $K^0_SK^0_S$ invariant-mass spectrum from BESIII in the
$X(2370)\to K^0_SK^0_S\pi^0$ channel. (b) Fit to the same spectrum for
events in the $K^*(892)^0$ signal region, showing the data, the total fit, the $X(2370)$
signal and the polynomial background.}
\label{fig:spectro}
\end{figure}

One of the characteristic predictions of non-Abelian gauge theories is the existence of glueballs,
bound states whose dominant Fock component is purely gluonic~\cite{ref:vadacchino}. Lattice QCD predicts a rich spectrum of such states, with the
lightest pseudoscalar ($J^{PC}=0^{-+}$) glueball expected between $2.3$ and
$3.0~\mathrm{GeV}/c^2$~\cite{ref:guietal}. Radiative $J/\psi$ decays therefore provide a favorable environment in which to search for
such states. Final states containing pseudoscalar mesons, such as $\pi^+\pi^-\eta'$, are
particularly useful because they can be studied with high statistics and have well-defined
quantum numbers.

BESIII reports comprehensive measurements of the $X(2370)$, discovered in 2011 in
$J/\psi\to\gamma\pi^+\pi^-\eta'$~\cite{ref:x2370disc} and confirmed in
$J/\psi\to\gamma K\bar K\eta'$~\cite{ref:x2370kk}. In 2024, using 10 billion $J/\psi$ events, its spin-parity was determined for the first
time to be $J^{PC}=0^{-+}$ through a partial-wave analysis of
$J/\psi\to\gamma K^0_S K^0_S\eta'$~\cite{ref:x2370jpc}. Since then, new decay modes have
been observed, including $X(2370)\to K^0_SK^0_S\pi^0$,
$\pi^0\pi^0\eta$ and $a_0\pi^0$~\cite{ref:x2370obs}; the corresponding $K^0_SK^0_S$ and
$K^0_S\pi^0$ invariant-mass spectra are shown in Fig.~\ref{fig:spectro}(a).

BESIII reports the $X(2370)$ mass, $(2376.3 \pm 8.7\,^{+3.2}_{-4.3})~\mathrm{MeV}/c^2$,
and width, $(83 \pm 17\,^{+44}_{-6})~\mathrm{MeV}$, both consistent with lattice-QCD predictions
for the lightest $0^{-+}$ glueball, a high production rate in $J/\psi$ radiative decays with
$\mathcal{B}(J/\psi\to\gamma X(2370)) > 10^{-3}$, an $\eta_c$-like decay pattern (including
$\pi^+\pi^-\eta'$, $K^+K^-\eta'$, $\pi^+\pi^-\eta$, $K^+K^-\eta$ and $K^+K^-\pi$) with suppressed
radiative decays to $\omega$ and $\phi$, and no evidence for $X(2370)\to K^*(892)^0\bar K^0$,
shown in Fig.~\ref{fig:spectro}(b), with
$\mathcal{B}[J/\psi\to\gamma X(2370)] \times \mathcal{B}[X(2370)\to K^*(892)^0\bar K^0 + \mathrm{c.c.}] < 2.7\times 10^{-6}$
at 90\% confidence level~\cite{ref:x2370glue}, implying
$\mathcal{B}(X(2370)\to K^*(892)\bar K) < 1.6\%$. The suppression of this mode is required by
generalized $G$-parity for a $0^{-+}$ flavor singlet, making the $X(2370)$ the first
flavor-singlet light hadron observed above $1~\mathrm{GeV}/c^2$ and disfavoring a conventional
$q\bar q$ assignment. A dominant $0^{-+}$ glueball component is essential for a natural and
complete explanation of these properties --- the mass, spin-parity, high production rate in
$J/\psi$ radiative decays, $\eta_c$-like decay pattern, flavor-singlet nature, narrow partial
decay width, and $\omega/\phi$ suppression --- since they are all consistent with the features
of the lightest $0^{-+}$ glueball while other interpretations can hardly explain them
simultaneously. This supports the lightest $0^{-+}$ glueball as the dominant constituent of the
$X(2370)$~\cite{ref:x2370glue}. Establishing its gluonic content quantitatively, however, still
requires a description of possible glueball--$q\bar q$ mixing and further measurements.

\section{Outlook and summary}

The physics prospects for the $e^+e^-$ heavy-flavor factories remain strong. Belle~II has
entered the $\mathrm{ab}^{-1}$ era and is expected to accumulate several $\mathrm{ab}^{-1}$
by 2032, with an ultimate goal of $50~\mathrm{ab}^{-1}$~\cite{ref:belle2tdr}. 
BESIII will continue to collect data with
the upgraded BEPCII collider, with dedicated data samples targeting charmed baryons
such as $\Lambda_c$ and $\Xi_c$~\cite{ref:bes3future}.

In summary, the $e^+e^-$ flavor factories are pursuing a broad physics program that spans
precision flavor measurements, tests of the Standard Model, and hadron spectroscopy.
The current $R(D)$ and $R(D^*)$ averages remain in tension with the SM at the approximately
$2\sigma$ level under the quoted averages and theoretical inputs. The BESIII measurements of $|V_{cd}|$ and $|V_{cs}|$ have
reached sub-percent precision, although both remain limited by the lattice-QCD form factors.
Belle~II has reported the first measurement of $S(\pi^0\pi^0)$, exploiting the quantum
entanglement of the $B^0\bar B^0$ pair. In hadron spectroscopy, Belle~II and BESIII continue
to deepen our understanding of QCD, from the newly observed $D$-wave bottomonium states to
the accumulating evidence that the $X(2370)$ may be the lightest pseudoscalar glueball.



\begin{thebibliography}{99}\small

\bibitem{PDG2026}
F. Takahashi et al. (Particle Data Group),
Review of Particle Physics,
Int. J. Mod. Phys. A 41, 2630011 (2026).

\bibitem{Altmannshofer:2025}
W.~Altmannshofer and P.~Stangl,
``Flavour Physics Beyond the Standard Model,''
[arXiv:2508.03950 [hep-ph]].

\bibitem{ref:belle2tdr}
T.~Abe et al.\ (Belle~II Collaboration),
\emph{Belle~II technical design report}, arXiv:1011.0352 [hep-ex].

\bibitem{ref:bes3future}
BESIII Collaboration,
\emph{Future physics programme of BESIII},
Chin.\ Phys.\ C \textbf{44}, 040001 (2020) [arXiv:1912.05983].

\bibitem{ref:knuunu}
Belle~II Collaboration,
\emph{Evidence for $B^+\to K^+\nu\bar\nu$},
Phys.\ Rev.\ D \textbf{109}, 112006 (2024).

\bibitem{ref:wilson}
\emph{Reinterpretation of $B^+\to K^+\nu\bar\nu$ in terms of Wilson coefficients},
Phys.\ Rev.\ D \textbf{112}, 092016 (2025).

\bibitem{ref:xsnn}
\emph{Search for $B\to X_s\nu\bar\nu$ with a fully reconstructed $X_s$},
Phys.\ Rev.\ Lett.\ \textbf{136}, 231801 (2026).

\bibitem{ref:ktatau}
Belle and Belle~II Collaborations,
\emph{Search for the $B^0\to K^0_S\tau^+\tau^-$ decay},
arXiv:2607.25607 [hep-ex].

\bibitem{ref:kptautau}
Belle and Belle~II Collaborations,
\emph{Search for the decay $B^+\to K^+\tau^+\tau^-$ using data from the Belle and Belle~II
experiments},
Phys.\ Rev.\ Lett.\ \textbf{137}, 051805 (2026).

\bibitem{ref:hpqcd}
HPQCD Collaboration,
\emph{Standard Model predictions for $B\to K\ell^+\ell^-$, $B\to K\ell^-_1\ell^+_2$ and
$B\to K\nu\bar\nu$ using form factors from $N_f=2+1+1$ lattice QCD},
Phys.\ Rev.\ D \textbf{107}, 014511 (2023) [arXiv:2207.12468].
[Erratum: Phys.\ Rev.\ D \textbf{107}, 119903 (2023)].

\bibitem{ref:fajfersm}
S.~Fajfer, J.~F.~Kamenik and I.~Nisandzic,
\emph{On the $B\to D^*\tau\bar\nu_\tau$ sensitivity to new physics},
Phys.\ Rev.\ D \textbf{85}, 094025 (2012) [arXiv:1203.2654].

\bibitem{ref:bernlochner}
F.~U.~Bernlochner, Z.~Ligeti, M.~Papucci and D.~J.~Robinson,
\emph{Tensions and correlations in $|V_{cb}|$ determinations},
Phys.\ Rev.\ D \textbf{95}, 115008 (2017) [arXiv:1703.05330].

\bibitem{ref:bigi}
D.~Bigi, P.~Gambino and S.~Schacht,
\emph{A fresh look at the determination of $|V_{cb}|$ from $B\to D^*\ell\nu$},
Phys.\ Lett.\ B \textbf{769}, 441 (2017) [arXiv:1703.06124].

\bibitem{ref:hflav}
HFLAV Collaboration,
\emph{Averages of $b$-hadron, $c$-hadron and $\tau$-lepton properties as of 2024},
Phys.\ Rev.\ D \textbf{113}, 012008 (2026) [arXiv:2411.18639].

\bibitem{ref:km}
M.~Kobayashi and T.~Maskawa,
\emph{CP violation in the renormalizable theory of weak interaction},
Prog.\ Theor.\ Phys.\ \textbf{49}, 652 (1973).

\bibitem{ref:tagvertex}
M.~Dorigo, S.~Raiz and D.~Tonelli,
\emph{Unlocking time-dependent CP violation without signal vertexing at $B$ factories},
Phys.\ Rev.\ D \textbf{112}, 032011 (2025) [arXiv:2506.11196].

\bibitem{ref:gnn}
Belle~II Collaboration,
\emph{New graph-neural-network flavor tagger for Belle~II and measurement of $\sin 2\phi_1$
in $B^0\to J/\psi K^0_S$ decays},
Phys.\ Rev.\ D \textbf{110}, 012001 (2024).

\bibitem{ref:spp}
Belle~II Collaboration,
\emph{Measurement of mixing-induced CP violation in the decay $B^0\to\pi^0\pi^0$},
arXiv:2606.20797 [hep-ex], submitted to Nature Physics.

\bibitem{ref:ckmfitter}
J.~Charles et al.\ (CKMfitter Group),
\emph{CP violation and the CKM matrix: assessing the impact of the asymmetric $B$
factories},
Phys.\ Rev.\ D \textbf{84}, 033005 (2011) [arXiv:1106.4041].

\bibitem{ref:vcs}
BESIII Collaboration,
\emph{Precise measurements of $D^0\to K^-\ell^+\nu_\ell$ and $D^+\to\bar K^0\ell^+\nu_\ell$
decays}, arXiv:2601.21196 [hep-ex].

\bibitem{ref:scalark}
BESIII Collaboration,
\emph{First experimental constraint on the scalar current in the
$D^{0(+)}\to\bar K\ell^+\nu_\ell$ transition}, arXiv:2601.21185 [hep-ex].

\bibitem{ref:vcd}
BESIII Collaboration,
\emph{Precision measurements of semileptonic decays $D^0\to\pi^-\ell^+\nu_\ell$ and
$D^+\to\pi^0\ell^+\nu_\ell$}, arXiv:2607.23945 [hep-ex].

\bibitem{ref:decdyn}
BESIII Collaboration,
\emph{Precision measurement of decay dynamics in $D^{0(+)}\to\pi^{-(0)}\ell^+\nu_\ell$},
arXiv:2607.23949 [hep-ex].

\bibitem{ref:yd}
Belle and Belle~II Collaborations,
\emph{First measurement of the masses of the $\Upsilon_1(1D)$ and $\Upsilon_3(1D)$ states
and the energy dependence of the cross sections for $e^+e^-\to\Upsilon_J(1D)\eta$ and
$e^+e^-\to\Upsilon_J(1D)\pi^+\pi^-$}, arXiv:2607.04369 [hep-ex], submitted to Phys.\ Rev.\ Lett.

\bibitem{ref:vadacchino}
D.~Vadacchino,
\emph{A review on glueball hunting},
arXiv:2305.04869 [hep-lat].

\bibitem{ref:guietal}
L.-C.~Gui, J.-M.~Dong, Y.~Chen and Y.-B.~Yang,
\emph{Study of the pseudoscalar glueball in $J/\psi$ radiative decays},
Phys.\ Rev.\ D \textbf{100}, 054511 (2019) [arXiv:1906.03666].

\bibitem{ref:x2370disc}
BESIII Collaboration,
\emph{Confirmation of the $X(1835)$ and observation of the resonances $X(2120)$ and
$X(2370)$ in $J/\psi\to\gamma\pi^+\pi^-\eta'$},
Phys.\ Rev.\ Lett.\ \textbf{106}, 072002 (2011) [arXiv:1012.3510].

\bibitem{ref:x2370kk}
BESIII Collaboration,
\emph{Observation of $X(2370)$ and search for $X(2120)$ in $J/\psi\to\gamma K\bar K\eta'$},
Eur.\ Phys.\ J.\ C \textbf{80}, 746 (2020) [arXiv:1912.11253].

\bibitem{ref:x2370jpc}
BESIII Collaboration,
\emph{Determination of spin-parity quantum numbers of $X(2370)$ as $0^{-+}$ from
$J/\psi\to\gamma K^0_SK^0_S\eta'$},
Phys.\ Rev.\ Lett.\ \textbf{132}, 181901 (2024) [arXiv:2312.05324].

\bibitem{ref:x2370obs}
BESIII Collaboration,
\emph{Observation of the $X(2370)$ in $J/\psi\to\gamma K^0_SK^0_S\pi^0$ and
$J/\psi\to\gamma\pi^0\pi^0\eta$}, arXiv:2605.26495 [hep-ex].

\bibitem{ref:x2370glue}
BESIII Collaboration,
\emph{Lightest $0^{-+}$ glueball as dominant constituent of the $X(2370)$},
arXiv:2607.20366 [hep-ex].

\end{thebibliography}
\end{document}